\documentclass[aps,pre,twocolumn,nofootinbib,floatfix]{revtex4}
\usepackage{amsmath}
\usepackage{epsfig}

\input{tcilatex}

\begin{document}

\title{Shot noise of quantum ring excitons in a planar microcavity}
\author{Y. N. Chen, D. S. Chuu, and S. J. Cheng}
\affiliation{Department of Electrophysics, National Chiao-Tung University, Hsinchu 300,
Taiwan}
\date{\today }

\begin{abstract}
Shot noise of quantum ring (QR) excitons in a p-i-n junction surrounded by a
microcavity is investigated theoretically. Some radiative decay properties
of a QR exciton in a microcavity can be obtained from the observation of the
current noise, which also gives the extra information about the tunnel
barriers. Different noise feature between the quantum dot (QD) and QR is
pointed out, and may be observed in a suitably designed experiment.

\end{abstract}

\maketitle

\address{Department of Electrophysics, National Chiao Tung University,
Hsinchu 300, Taiwan}





Since Purcell proposed the idea of controlling the spontaneous emission rate
by using a cavity \cite{1}, the enhanced and inhibited SE rate for the
atomic system was intensively investigated in the 1980s \cite{2} by using
atoms passed through a cavity. In semiconductor systems,\ the electron-hole
pair is naturally a candidate to examine the spontaneous emission.
Modifications of the SE rates of the QD \cite{3}, quantum wire \cite{4}, or
quantum well \cite{5} excitons inside the microcavities have been observed
experimentally.

Recently, interest in measurements of shot noise spectrum has risen owing to
the possibility of extracting valuable information not available in
conventional dc transport experiments \cite{6}. With the advances of
fabrication technologies, it is now possible to embed the QDs inside a p-i-n
structure \cite{7}, such that the electron and hole can be injected
separately from the opposite sides. This allows one to examine the exciton
dynamics in a QD via electrical currents. \cite{8} On the other hand, it is
now possible to fabricate the ring-shaped dots of InAs in GaAs with a
circumference of several hundred nanometers \cite{9}. Optical detection of
the Aharonov-Bohm effect on an exciton in a single QR was also reported \cite%
{10}.

Based on the rapid progress of nano-technologies, it's not hard to imagine
that the QR can be incorporated in a p-i-n junction surrounded by the
microcavity. Examinations of the dynamics of the QR excitons by the
electrical currents can soon be realized. We thus present in this work the
non-equilibrium calculations of such a device. Current noise of QR excitons
in a planar microcavity is obtained via the MacDonald formula \cite{11}, and
is found to reveal some characteristics of the restricted environment, i.e.
the density of states of the confined photons. 
\begin{figure}[th]
\includegraphics[width=7.5cm]{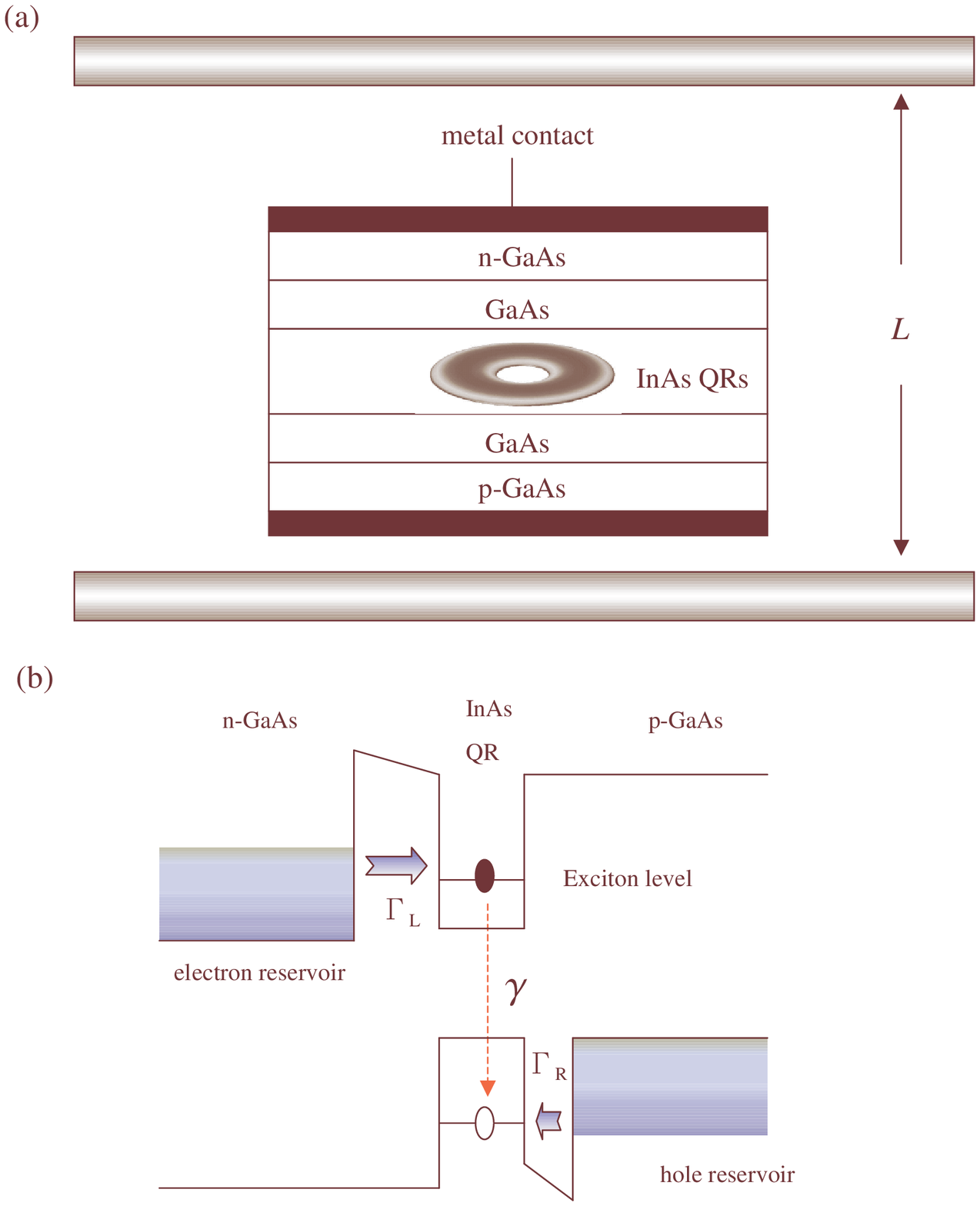}
\caption{{}(Color online). (a) Schematic description of a QR inside a p-i-n
junction surrounded by a planar microcavity with length $L$. (b) Energy-band
diagram of a QR in the p-i-n junction.}
\end{figure}

\emph{The model.} --- Consider now a QR embedded in a p-i-n junction as
shown in Fig. 1. Both the hole and electron reservoirs are assumed to be in
thermal equilibrium. For the physical phenomena we are interested in, the
fermi level of the \textit{p(n)}-side hole (electron) is slightly lower
(higher) than the hole (electron) subband in the dot. After a hole is
injected into the hole subband in the QR, the \textit{n}-side electron can
tunnel into the exciton level because of the Coulomb interaction between the
electron and hole. Thus, we may introduce the three ring states: $\left|
0\right\rangle =\left| 0,h\right\rangle $, $\left| \uparrow \right\rangle
=\left| e,h\right\rangle $, and $\left| \downarrow \right\rangle =\left|
0,0\right\rangle $,where $\left| 0,h\right\rangle $ means there is one hole
in the QR,\ $\left| e,h\right\rangle $ is the exciton state, and $\left|
0,0\right\rangle $ represents the ground state with no hole and electron in
the QR. One might argue that one can not neglect the state $\left|
e,0\right\rangle $ for real device since the tunable variable is the applied
voltage. This can be resolved by fabricating a thicker barrier on the
electron side so that there is little chance for an electron to tunnel in
advance. Moreover, the charged exciton and biexcitons states are also
neglected in our calculations. This means a low injection limit is required
in the experiment\cite{7}.

\emph{Derivation of Master equation.} ---We can now define the
ring-operators $\overset{\wedge }{n_{\uparrow }}\equiv \left| \uparrow
\right\rangle \left\langle \uparrow \right| ,$ $\overset{\wedge }{%
n_{\downarrow }}\equiv \left| \downarrow \right\rangle \left\langle
\downarrow \right| ,$ $\overset{\wedge }{p}\equiv \left| \uparrow
\right\rangle \left\langle \downarrow \right| ,$ $\overset{\wedge }{%
s_{\uparrow }}\equiv \left| 0\right\rangle \left\langle \uparrow \right| ,$ $%
\overset{\wedge }{s_{\downarrow }}\equiv \left| 0\right\rangle \left\langle
\downarrow \right| $. The total Hamiltonian $H$ of the system consists of
three parts: $H_{0}$ [ring, photon bath $H_{p}$, and the electron (hole)
reservoirs $H_{res}$], $H_{T}$ (ring-photon coupling), and the
ring-reservoir coupling $H_{V}$:

\begin{eqnarray}
H &=&H_{0}+H_{T}+H_{V}  \notag \\
H_{0} &=&\varepsilon _{\uparrow }\overset{\wedge }{n_{\uparrow }}%
+\varepsilon _{\downarrow }\overset{\wedge }{n_{\downarrow }}+H_{p}+H_{res} 
\notag \\
H_{T} &=&\sum_{k}(D_{k}b_{k}^{\dagger }\overset{\wedge }{p}+D_{k}^{\ast
}b_{k}\overset{\wedge }{p}^{\dagger })=\overset{\wedge }{p}X+\overset{\wedge 
}{p}^{\dagger }X^{\dagger }  \notag \\
H_{p} &=&\sum_{k}\omega _{k}b_{k}^{\dagger }b_{k}  \notag \\
H_{V} &=&\sum_{\mathbf{q}}(V_{\mathbf{q}}c_{\mathbf{q}}^{\dagger }\overset{%
\wedge }{s_{\uparrow }}+W_{\mathbf{q}}d_{\mathbf{q}}^{\dagger }\overset{%
\wedge }{s_{\downarrow }}+c.c.)  \notag \\
H_{res} &=&\sum_{\mathbf{q}}\varepsilon _{\mathbf{q}}^{\uparrow }c_{\mathbf{q%
}}^{\dagger }c_{\mathbf{q}}+\sum_{\mathbf{q}}\varepsilon _{\mathbf{q}%
}^{\downarrow }d_{\mathbf{q}}^{\dagger }d_{\mathbf{q}}.
\end{eqnarray}%
In the above equation, $b_{k}$ is the photon operator, $D_{k}$ is the dipole
coupling strength, $X=\sum_{k}D_{k}b_{k}^{\dagger }$ ,$\ $and $c_{\mathbf{q}%
} $ and $d_{\mathbf{q}}$ denote the electron operators in the left ad right
reservoirs, respectively. The couplings to the electron and hole reservoirs
are given by the standard tunnel Hamiltonian $H_{V},$ where $V_{\mathbf{q}}$
and $W_{\mathbf{q}}$ couple the channels $\mathbf{q}$ of the electron and
the hole reservoirs. If the couplings to the electron and the hole
reservoirs are weak, then it is reasonable to assume that the standard
Born-Markov approximation with respect to these couplings is valid. In this
case, one can derive a master equation from the exact time-evolution of the
system. The equations of motion can be expressed as (cf. [12])

\begin{eqnarray}
\frac{\partial }{\partial t}\overset{\wedge }{\left\langle n_{\uparrow
}\right\rangle }_{t} &=&-\int dt^{\prime }[C(t-t^{\prime })+C^{\ast
}(t-t^{\prime })]\overset{\wedge }{\left\langle n_{\uparrow }\right\rangle }%
_{t^{\prime }}  \notag \\
&&+\Gamma _{L}[1-\overset{\wedge }{\left\langle n_{_{\uparrow
}}\right\rangle }_{t}-\overset{\wedge }{\left\langle n_{\downarrow
}\right\rangle }_{t}]
\end{eqnarray}

\begin{equation*}
\frac{\partial }{\partial t}\overset{\wedge }{\left\langle n_{\downarrow
}\right\rangle }_{t}=\int dt^{\prime }[C(t-t^{\prime })+C^{\ast
}(t-t^{\prime })]\overset{\wedge }{\left\langle n_{\uparrow }\right\rangle }%
_{t^{\prime }}-\Gamma _{R}\overset{\wedge }{\left\langle n_{\downarrow
}\right\rangle }_{t}]
\end{equation*}

\begin{equation*}
\frac{\partial }{\partial t}\overset{\wedge }{\left\langle p\right\rangle }%
_{t}=-\frac{1}{2}\int dt^{\prime }[C(t-t^{\prime })+C^{\ast }(t-t^{\prime })]%
\overset{\wedge }{\left\langle p\right\rangle }_{t^{\prime }}-\frac{\Gamma
_{R}}{2}\overset{\wedge }{\left\langle p\right\rangle }_{t},
\end{equation*}%
where $\Gamma _{L}$ $=2\pi \sum_{\mathbf{q}}V_{\mathbf{q}}^{2}\delta
(\varepsilon _{\uparrow }-\varepsilon _{\mathbf{q}}^{\uparrow })$ , $\Gamma
_{R}=2\pi \sum_{\mathbf{q}}W_{\mathbf{q}}^{2}\delta (\varepsilon
_{\downarrow }-\varepsilon _{\mathbf{q}}^{\downarrow })$, and $\varepsilon
=\varepsilon _{\uparrow }-\varepsilon _{\downarrow }$ is the energy gap of
the QR exciton. Here, $C(t-t^{\prime })$ $\equiv \left\langle
X_{t}X_{t^{\prime }}^{\dagger }\right\rangle _{0}$ is the photon correlation
function, and depends on the time interval only. We can now define the
Laplace transformation for real $z,$

\begin{eqnarray}
C_{\varepsilon }(z) &\equiv &\int_{0}^{\infty }dte^{-zt}e^{i\varepsilon
t}C(t)  \notag \\
n_{\uparrow }(z) &\equiv &\int_{0}^{\infty }dte^{-zt}\overset{\wedge }{%
\left\langle n_{\uparrow }\right\rangle }_{t}\text{ \ }etc.,\text{ }z>0
\end{eqnarray}%
and transform the whole equations of motion into $z$-space,

\begin{eqnarray}
n_{\uparrow }(z) &=&-(C_{\varepsilon }(z)+C_{\varepsilon }^{\ast
}(z))n_{\uparrow }(z)/z  \notag \\
&&+\frac{\Gamma _{L}}{z}(1/z-n_{\uparrow }(z)-n_{\downarrow }(z))  \notag \\
n_{\downarrow }(z) &=&(C_{\varepsilon }(z)+C_{\varepsilon }^{\ast
}(z))n_{\downarrow }(z)/z-\frac{\Gamma _{R}}{z}n_{\downarrow }(z)  \notag \\
p(z) &=&-\frac{1}{2}(C_{\varepsilon }(z)+C_{\varepsilon }^{\ast }(z))p(z)/z-%
\frac{\Gamma _{R}}{2z}p(z).
\end{eqnarray}%
These equations can then be solved algebraically, and the tunnel current
from the hole- or electron-side barrier

\begin{equation}
\overset{\wedge }{I}_{R}=-e\Gamma _{R}\overset{\wedge }{\left\langle
n_{\downarrow }\right\rangle }_{t},\text{ \ }\overset{\wedge }{I}%
_{L}=-e\Gamma _{L}[1-\overset{\wedge }{\left\langle n_{\uparrow
}\right\rangle }_{t}-\overset{\wedge }{\left\langle n_{\downarrow
}\right\rangle }_{t}]
\end{equation}
can then be obtained by performing the inverse Laplace transformation on
Eqs. (4).

\emph{Shot noise spectrum.} --- In a quantum conductor in nonequilibrium,
electronic current noise originates from the dynamical fluctuations of the
current being away from its average. To study correlations between carriers,
we relate the exciton dynamics with the hole reservoir operators by
introducing the degree of freedom $n$ as the number of holes that have
tunneled through the hole-side barrier \cite{13} and write 
\begin{gather}
\overset{\cdot }{n}_{0}^{(n)}(t)=-\Gamma _{L}n_{0}^{(n)}(t)+\Gamma
_{R}n_{\downarrow }^{(n-1)}(t),  \notag \\
\overset{\cdot }{n}_{\uparrow }^{(n)}(t)+\overset{\cdot }{n}_{\downarrow
}^{(n)}(t)=(\Gamma _{L}-\Gamma _{R})n_{0}^{(n)}(t).
\end{gather}%
Eqs. (6) allow us to calculate the particle current and the noise spectrum
from $P_{n}(t)=n_{0}^{(n)}(t)+n_{\uparrow }^{(n)}(t)+n_{\downarrow
}^{(n)}(t) $ which gives the total probability of finding $n$ electrons in
the collector by time $t$. In particular, the noise spectrum $S_{I_{R}}$ can
be calculated via the MacDonald formula \cite{14}. In the zero-frequency
limit, the Fano factor can be written as

\begin{eqnarray}
F &\equiv &\frac{S_{I_{R}}(\omega =0)}{2e\langle I\rangle } \\
&=&1-\left. \frac{A(z)\Gamma _{L}\Gamma _{R}[A(z)+2(\Gamma _{L}+\Gamma _{R})]%
}{\{A(z)\Gamma _{R}+\Gamma _{L}[A(z=0)+2\Gamma _{R}]\}^{2}}\right| _{z=0}, 
\notag
\end{eqnarray}%
where $A(z)\equiv C_{\varepsilon }(z)+C_{\varepsilon }^{\ast }(z)$.

\emph{Results and Discussions.} --- From Eq. (7), one knows that the noise
spectrum of the QR excitons depends strongly on $C_{\varepsilon }(z)$, which
reduces to the radiative decay rate $\gamma $ in the Markovian limit. The
exciton decay rate $\gamma $ in a microcavity can be obtained easily from
the perturbation theory and is given by

\begin{eqnarray}
\gamma &=&\frac{e^{2}\hbar }{m^{2}c}\frac{\rho }{d}\int \left|
H_{0}^{(1)}(q^{\prime }\rho )\right| ^{2}q^{\prime }  \notag \\
&&(\sum_{k_{z}^{\prime }}\frac{\delta (\varepsilon -c\sqrt{q^{\prime
2}+k_{z}^{\prime 2}})}{\sqrt{k_{z}^{\prime 2}+q^{\prime 2}}}\left| \mathbf{%
\epsilon }_{\mathbf{q}^{\prime }k_{z}^{\prime }\lambda }\cdot \mathbf{\chi }%
\right| ^{2})dq^{\prime },
\end{eqnarray}%
where $\rho $ is the ring radius, $d$ is the lattice spacing, $H_{0}^{(1)}$
is the Hankel function, $\mathbf{\epsilon }_{\mathbf{q}^{\prime
}k_{z}^{\prime }}$ is the polarization of the photon, and $\mathbf{\chi }$
is the dipole moment of the QR exciton. \cite{15} The summation of the
integer modes in the $k_{z}^{\prime }$ direction is determined from the
boundary conditions of the planar microcavity. 
\begin{figure}[th]
\includegraphics[width=7.5cm]{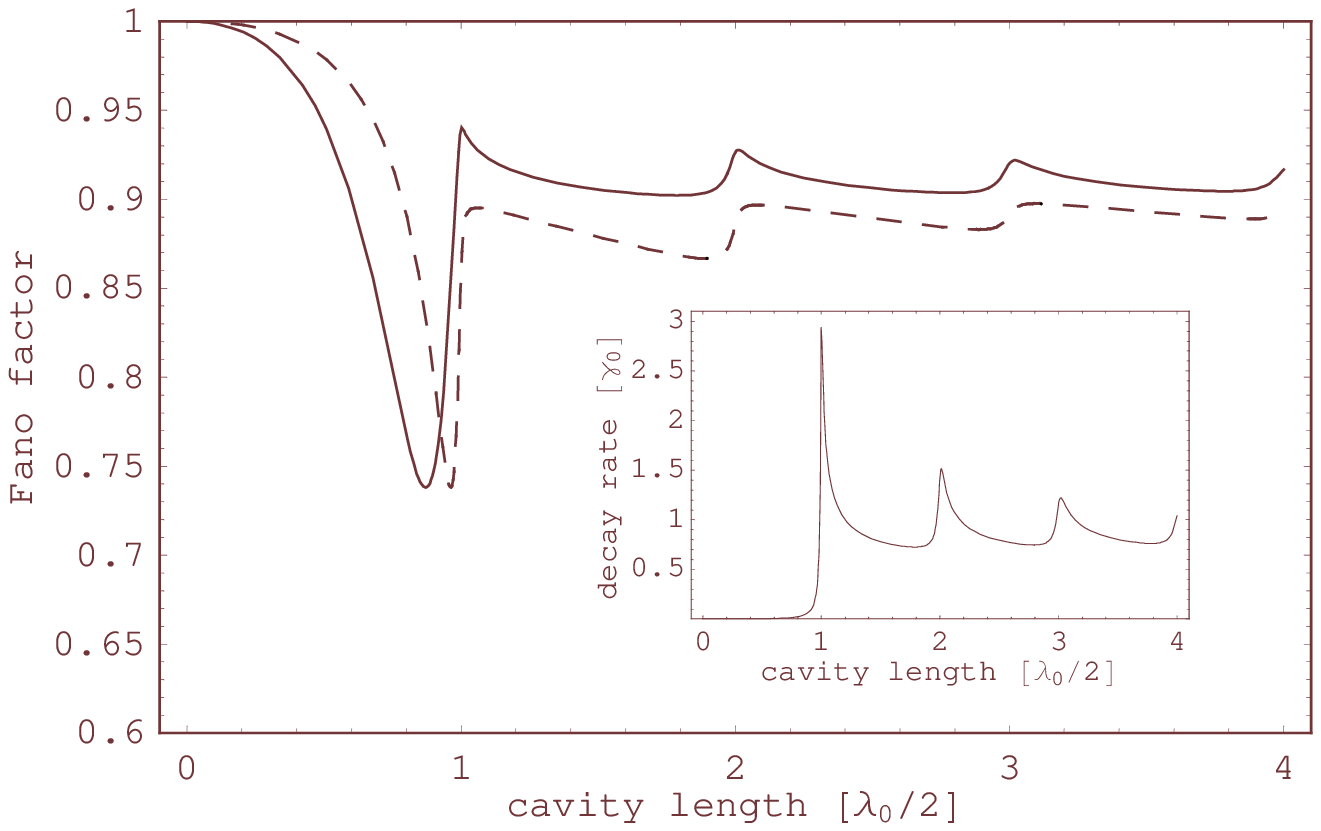}
\caption{{}Fano factor as a function of cavity length $L$. The vertical and
horizontal units are $\frac{S_{I_{R}}(0)}{2eI}$ and $\protect\lambda _{0}/2$%
, respectively. The inset shows the radiative decay rate of a QR exciton in
a planar microcavity. }
\end{figure}

The radiative decay rate $\gamma $ of a QR exciton inside a planar
microcavity is numerically displayed in the inset of Fig. 2. The tunneling
rates, $\Gamma _{L}$ and $\Gamma _{R}$, are assumed to be equal to 0.1$%
\gamma _{0}$ and $\gamma _{0}$, where $\gamma _{0}$ ($\sim $1/1ns) is the
decay rate of a QR exciton in free space. Also, the planar microcavity is
assumed to have a Lorentzian broadening at each resonant mode (with
broadening width equals to 1\% of each resonant mode). \cite{16} As the
cavity length is shorter than one half the wavelength of the emitted photon (%
$L<\lambda _{0}/2$), the decay rate is inhibited because of the cut-off
frequency of the cavity. When the cavity length exceeds some multiple
wavelength, it opens up another decay channel for the quantum ring exciton
and turns out that there is an abrupt enhancement on the decay rate. Such a
singular behavior also happens in the decay of one dimensional quantum wire
polaritons inside a microcavity. \cite{16} This is because the ring geometry
preserves the angular momentum of the exciton, rendering the formation of
exciton-polariton in the direction of circumference. This kind of behavior
can also be found in the calculations of Fano factor as demonstrated by the
solid line in Fig. 2. Comparing to the zero-frequency noise of the QD
excitons (dashed line), the Fano factor of the QR excitons shows the
''cusp'' feature at each resonant mode.

\begin{figure}[th]
\includegraphics[width=7.5cm]{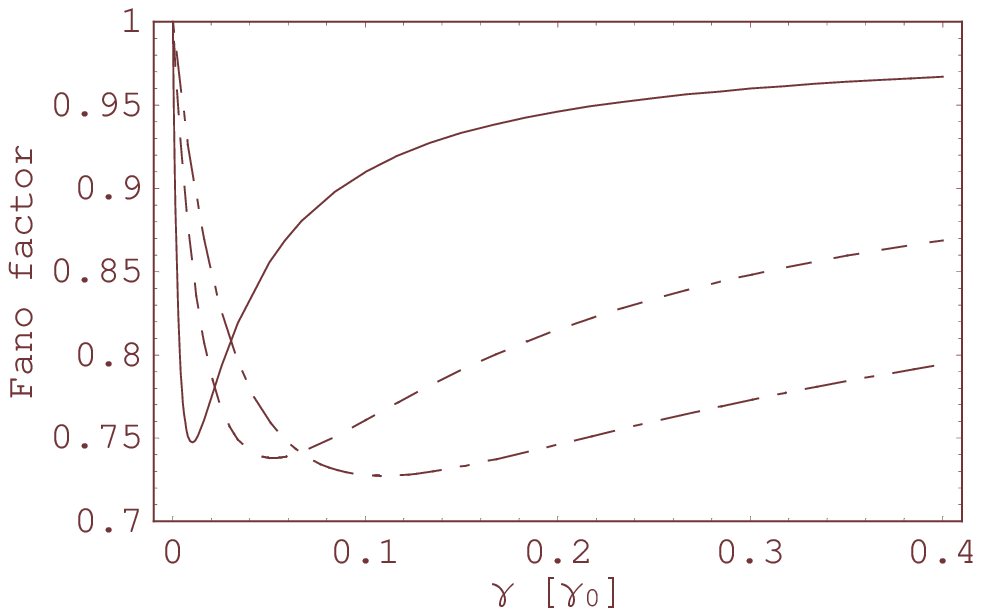}
\caption{{}Fano factor as a function of the decay rate for different
electron-side tunneling rate: $\Gamma _{L}=0.01\protect\gamma _{0}$ (solid
line), $0.05\protect\gamma _{0}$ (dashed line), and $0.1\protect\gamma _{0}$
(dashed-doted line). The hole-side tunneling rate is fixed ($\Gamma _{R}=%
\protect\gamma _{0}$). }
\end{figure}

Another interesting point is that below the lowest resonant mode ($L=\lambda
_{0}/2$), both the solid and dashed curves have a dip in the Fano factor. It
is not seen from the radiative decay rate. To answer this, we have plotted
Eq. (7) in Fig. 3 as a function of the decay rate $\gamma $. Keeping $\Gamma
_{R}$ unchanged, the solid, dashed, and dashed-dotted lines correspond to
the electron-side tunneling rate $\Gamma _{L}=0.01\gamma _{0},0.5\gamma _{0}$%
, and $0.1\gamma _{0}$, respectively. As can be seen, the Fano factor has a
minimum point at $\gamma =$ $\Gamma _{L}\Gamma _{R}(\Gamma _{L}+\Gamma
_{R})/(\Gamma _{L}^{2}+\Gamma _{R}^{2})$. Comparing this with the inset of
Fig. 2, one immediately knows that when the cavity length is increased to $%
\lambda _{0}/2$, the abruptly increased decay rate will cross the minimum
point and result in a dip in Fig. 2. Furthermore, in the limit of $\Gamma
_{R}>>\Gamma _{L}$, the minimum point can be approximated as: $\gamma
\approx \Gamma _{L}$ . This means by observing the dip in the Fano factor of
Fig. 2, the magnitude of the electron-side tunneling rate $\Gamma _{L}$ can
be obtained. 
\begin{figure}[th]
\includegraphics[width=7.5cm]{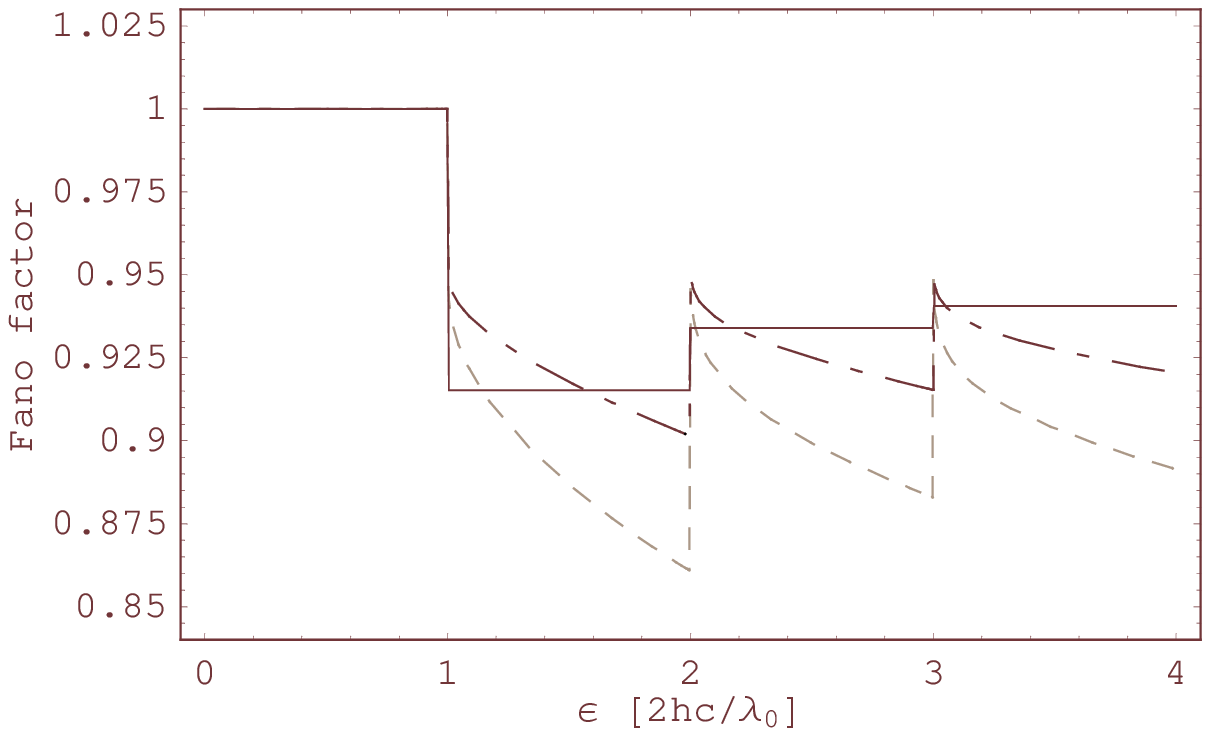}
\caption{(Color{} online). Fano factor of QD (solid line) and QR (gray
dashed line) excitons as a function of energy gap $\protect\varepsilon $.
The vertical and horizontal units are $\frac{S_{I_{R}}(0)}{2eI}$ and $2hc/%
\protect\lambda _{0}$, respectively. The dashed-dotted line represents that
if the free-space decay rate of the QR exciton is enhaced by a factor of 2 ($%
\protect\gamma _{0}\rightarrow 2\protect\gamma _{0}$), the shot noise is
also enhanced.}
\end{figure}

To further understand the difference between the QD and QR excitons, Fig. 4
illustrates the shot noise as a function of energy gap $\varepsilon $. In
plotting the figure, a perfect planar microcavity is assumed for
convenience. As can be seen, the shot noise of the QD excitons shows the
plateau feature (solid line) with the increasing of $\varepsilon $, while it
is a zigzag behavior (gray-dashed and dashed-dotted lines) for the QR
excitons. This is because the decay rate of the QD exciton in a microcavity
is given by \cite{16} 
\begin{equation}
\gamma _{dot}\propto \sum_{n_{c}}\frac{e^{2}\hbar }{m^{2}c^{2}L}\theta
((\varepsilon /\hbar c)^{2}-(\pi n_{c}/L)^{2})\left| \mathbf{\epsilon }_{%
\mathbf{q}^{\prime }k_{z}^{\prime }\lambda }\cdot \mathbf{\chi }\right| ^{2},
\end{equation}%
where $\theta $ is the step function, and the summation is over the positive
integers. Therefore, when the energy gap $\varepsilon $ is tuned above some
resonant mode of the cavity, the decay rate is a constant before the next
decay channel is opened. On the other hand, however, one knows that the
decay rate for the QR exciton is not a constant between two resonant modes.
This explains why the decay property for QR exciton is different from that
for the QD exciton under the same photonic environment, and the difference
may be distinguished by the shot noise measurements. From the experimental
point of view, different dependences on $\varepsilon $ are easier to be
realized since it's almost impossible to vary the cavity length once the
sample is prepared. A possible way to observe the mentioned effects is to
vary $\varepsilon $ around the discontinuous points and measure the
corresponding current noise.

A few remarks about the ring radius should be mentioned here. One should
note that we do not give the specific value of the ring radius in our model.
Instead, a phenomenological value about the free-space decay rate $\gamma
_{0}$ is used. The magnitudes of the tunnel rates are set relative to it. In
general, the changing of radius will certainly affect the shot noise. For
example, because of the exciton-polariton (superradiant) effect in the
direction of circumference, an increasing of ring radius will enhance the
decay rate. In addition, the dipole moment of the QR exciton $\chi $ is also
altered because of the varying of the wavefunction. All these can contribute
to the variations of the decay rate and shot noise. In our previous study %
\cite{15}, we have shown that the decay rate is a monotonic increasing
function on radius $\rho $ if the exciton is coherent in the quantum ring,
i.e. free of scattering from impurities or imperfect boundaries. The
dashed-dotted line in Fig. 4 shows the result for \emph{doubled} free-space
decay rate, i.e. $\gamma _{0}\rightarrow 2\gamma _{0}$. Although the noise
is increased, the zigzag feature remains unchanged.

In conclusion, we have derived in this work the non-equilibrium current
noise of QR excitons incorporated in a p-i-n junction surrounded by a planar
microcavity. Some radiative decay properties of the one-dimensional QR
exciton can be obtained from the observation of shot noise spectrum, which
also shows extra information about the electron-side tunneling rate.
Different noise feature between the QD and QR is pointed out, and deserved
to be tested with present technologies.

This work is supported partially by the National Science Council, Taiwan
under the grant numbers NSC 94-2112-M-009 -019 and NSC 94-2120-M-009-002.


\begin{thebibliography}{99}
\bibitem{13} Actually, the total current noise should be expressed in terms
of the spectra of particle currents and the charge noise spectrum: $%
S_{I}(\omega )=aS_{I_{L}}(\omega )+bS_{I_{R}}(\omega )-ab\omega
^{2}S_{Q}(\omega )$, where $a$ and $b$ are capacitance coefficient ($a+b=1$)
of the junctions. Since we have assumed a highly asymmetric set-up ($a<<b$),
it is plausible to consider the hole-side spectra $S_{I_{R}}(\omega )$ only.
\end{thebibliography}
\end{document}